\documentclass[letterpaper, 10pt, conference]{ieeeconf}
\IEEEoverridecommandlockouts
\usepackage{amssymb}
\usepackage{amsmath}
\usepackage{array}
\usepackage{cite}
\usepackage{graphicx}
\usepackage{hyperref}

\usepackage{xcolor}

\title{\LARGE \bf Human Balancing on a Log: A Switched Multi-Layer Controller}

\author{Jiayi Zhao, Mo Yang, Jing Shuang (Lisa) Li
	\thanks{$^*$All authors were affiliated with the Department of Electrical Engineering and Computer Science at the University of Michigan at the time this paper was written. Corresponding author:
 {\tt\small jslisali@umich.edu}}
}

\begin{document}

\maketitle

\begin{abstract}
We study the task of balancing a human on a log that is fixed in place. Balancing on a log is substantially more challenging than balancing on a flat surface due to increased instability --- nonetheless, we are able to balance by composing simple (e.g., PID, LQR) controllers in a bio-inspired switched multi-layer configuration. The controller consists of an upper-layer LQR planner (akin to the central nervous system) that coordinates ankle and hip torques, and lower-layer PID trackers (akin to local motor units) that follow this plan subject to nonlinear dynamics. The controller switches between three operational modes depending on the state of the human. The efficacy of the controller is verified in simulation, where our controller is able to stabilize the human for a variety of initial conditions and disturbances. We also introduce a controller that outputs muscle activations to perform the same balancing task.
\end{abstract}

\section{Introduction}

Postural balancing is an important part of everyday life --- most humans stand and balance with ease. Many models of postural balancing have been developed \cite{Ivanenko2018-do, kanamiya2010ankle, Muhsen, Barton2016-hm, Kuo362914, Kamran6091931}, which improve our scientific understanding of motor control and propose control mechanisms for postural balance in robots. Some models make simplifying assumptions that result in an inverted pendulum model \cite{Muhsen, Kuo362914, Kamran6091931}, while others incorporate more biomechanical details \cite{Barton2016-hm}; optimal control \cite{Kuo362914, Kamran6091931} and switched control \cite{kanamiya2010ankle} strategies are often used.

Typically, models of postural balancing involve balancing on flat ground \cite{Muhsen, Barton2016-hm, Kuo362914, Kamran6091931}; however, humans are also capable of balancing on more difficult surfaces (e.g., slanted surfaces). In this work, we are interested in the problem of balancing on a cylinder (i.e., ``log") that is fixed in place, as shown in Figure \ref{fig:model_main}. This task features highly unstable nonlinear dynamics, in which even small deviations from equilibrium result in rapidly growing accelerations. Nonetheless, most humans with sufficient fitness and coordination can maintain balance for at least several seconds without prior training\footnote{We invite the reader to try this at home or outdoors.}. Thus, \textbf{the goal of this paper is to provide a simple control strategy that allows a humanoid to balance on a log}. To the best of our knowledge, this task has not been studied. Existing models for balancing on a flat ground do not easily extend to this scenario, as they typically treat the problem as a single- or double-inverted pendulum with a fixed base; however, in our case, the contact point between the foot and the log is not fixed, necessitating a different control approach with more reliance on torque from the hip joint.

We choose a controller that is inspired by animal sensorimotor control, which features multiple layers. Upper layers (e.g. central nervous system) are responsible for planning and coordination across body parts, while lower layers (e.g. motor units associated with individual joints and muscles) are responsible for carrying out these plans for individual body parts \cite{Nakahira2021diversity, Karashchuk2024}. In addition to being multi-layered, our controller also switches between different operational modes depending on the current state of the human, leveraging ideas from switched strategies in postural control \cite{kanamiya2010ankle}. The overall controller consists of an upper-layer linear quadratic regulator (LQR) and lower-layer proportional-integral-derivative (PID) controller (Section \ref{sec:methodology}); \textbf{by composing these simple controllers in a layered, switched configuration, we are able to achieve stability for a variety of initial conditions and disturbances} (Section \ref{sec:simulations}). We also briefly describe how this overall model can be made more biologically plausible by incorporating muscle-based actuation (Section \ref{sec:muscle_controller}), and discuss future work (Section \ref{sec:conclusions}).

\section{Problem setup} \label{sec:problem_setup}
\begin{figure}
    \centering \includegraphics[width=.9\hsize, page=1]{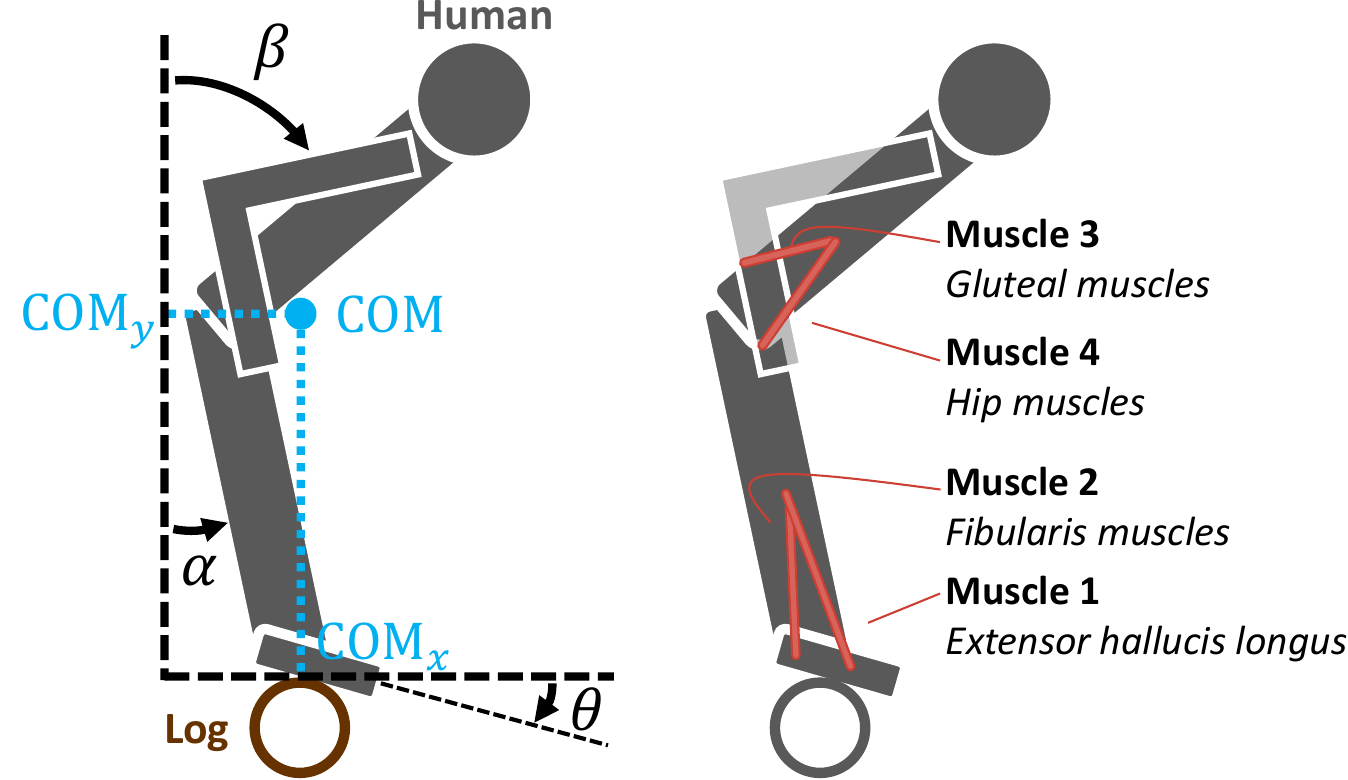}
    \caption{\textbf{(Left)} Human balancing on a log. We use right-hand sign convention for angles; here, $\beta$ and $\theta$ have negative values. \textbf{(Right)} Schematic of muscle-based model, described in Section \ref{sec:muscle_controller}. Here, instead of applying torques to the ankle and hip joints, we manipulate joints using simplified muscles.}
    \label{fig:model_main}
\end{figure}

We consider a human balancing on a log that is fixed in place and does not roll, as depicted in Figure \ref{fig:model_main}. The contact point between the foot and the log is variable, and we assume that the foot does not slip on the log. Analysis is confined to the depicted 2-D plane. In this model, the body contains three mass points, located at the head, hip and ankle. The knees are assumed to be straight, and there is no hand or arm motion; the hip and ankle are free to rotate. The angles of interest are $\theta$, $\alpha$, and $\beta$. $\theta$ indicates angular deviation of the foot from the $x$ axis, and $\alpha$ and $\beta$ indicate angular deviations from the $y$ axis for the leg and torso, respectively. Note that these are global angles, which can be converted to body angles; specifically, hip angle can be expressed as $\pi - \alpha + \beta$, and ankle angle can be expressed as $0.5\pi + \alpha -\theta$. 

The nonlinear equations of motion for the human balancing on a log are given in \eqref{eq:eom_torque}. Here, $x_i$ and $y_i$ are the projections of the $i^{\text{th}}$ mass point on the $x$ and $y$ axes, respectively. $r$ is the radius of the log; $l_i$ are the distances from the $i^{\text{th}}$ mass to the $(i-1)^{\text{th}}$ mass; $l_0$ is the distance of the 1st mass point to the middle of the foot. $\text{COM}_x$ and $\text{COM}_y$ are the projections of the center of mass (COM) on the $x$ and $y$ axes; $F_x$ and $F_y$ are forces along the $x$ and $y$ axes; $N_x$ and $N_y$ are the normal forces from the log to the foot projected along the $x$ and $y$ axes. $m_i$ are the masses of the mass points. $\tau_1$ is ankle torque, and $\tau_2$ is hip torque.

The goal of the controller is to restore the human to the equilibrium posture: the foot is parallel with the ground ($\theta = 0$), the leg is vertical to the ground ($\alpha = 0$), and the torso is slightly leaning forward ($\beta = \beta_0$ for some small $\beta_0$), as shown in the top of Figure \ref{fig:diagram_cases} (``Case 1"). This corresponds to $\text{COM}_x=0$. Define $\tilde{\beta} := \beta - \beta_0$. The controller has access to angles $(\theta, \alpha, \tilde{\beta})$, angular velocities $(\dot{\theta}, \dot{\alpha}, \dot{\tilde{\beta}})$, and $\text{COM}_x$\footnote{Note: only $\text{COM}_x$ is relevant to stability; $\text{COM}_y$ can be ignored.}, and outputs ankle torque $\tau_1$ and hip torque $\tau_2$.

\begin{figure}
\begin{equation} \label{eq:eom_torque}
    \begin{aligned}
        x_0 &= -r\sin\theta + \theta r \cos\theta - l_0\cos\theta \\
        x_1 &= x_0 - l_1 \sin\alpha, \quad x_2 = x_1 - l_2 \sin\beta \\
        \text{COM}_x &= \frac{x_0 \, m_0 + x_1 \, m_1 + x_2 \, m_2}{m_0 + m_1 + m_2} \\
        N_x &=  \frac{\tau_1}{r\theta-l_0} \sin\theta \\
        F_x + N_x &= \ddot{x}_0 \, m_0 + \ddot{x}_1 \, m_1 + \ddot{x}_2 \, m_2 \\
        y_0 &= r \cos\theta  + \theta r \sin\theta  - l_0 \sin\theta \\
        y_1 &= y_0 + l_1 \cos\alpha, \quad y_2 = y_1 + l_2 \cos\beta \\
        \text{COM}_y &= \frac{y_0 \, m_0 + y_1 \, m_1 + y_2 \, m_2}{m_0 + m_1 + m_2} \\
        N_y &= \frac{\tau_1}{l_0 - r\theta} \cos\theta \\
        F_y + N_y &= (\ddot{y}_0 + g) \, m_0 + (\ddot{y}_1 + g) \, m_1 + (\ddot{y}_2 + g) \, m_2 \\[10pt]
        \tau_1 &= - \ddot{x}_2 m_2(l_2 \cos\beta + l_1\cos\alpha ) \\
         & \quad - m_2 (\ddot{y}_2 + g) (l_2 \sin\beta  + l_1 \sin\alpha) \\
         & \quad - \ddot{x}_1 l_1 m_1 \cos\alpha - (\ddot{y}_1 + g) l_1 m_1 \sin\alpha \\
        \tau_2 &= - \ddot{x}_2 l_2 m_2\cos\beta - (\ddot{y}_2 + g) l_2 m_2 \sin\beta \\
        F_y &= F_x\tan\theta \\
        N_x^2 + N_y^2 &= N_x (\ddot{x}_0 \, m_0 + \ddot{x}_1 \, m_1 + \ddot{x}_2 \, m_2) \\ 
        & \quad + N_y [(\ddot{y}_0 + g) \, m_0 + (\ddot{y}_1 + g) \, m_1
        + (\ddot{y}_2 + g) \, m_2 ] \\
        0 &= (f_y + N_y) (\text{COM}_x - r \sin\theta) \\
        & \quad + (f_x + N_x) (\text{COM}_y - r \cos\theta) \\
        & \quad - \ddot{x}_0 \, m_0 (\text{COM}_y - y_0) - \ddot{x}_1 \, m_1 (\text{COM}_y - y_1) \\
        & \quad - \ddot{x}_2 \, m_2 (\text{COM}_y - y_2) - \ddot{y}_0 \, m_0 (x_0 - \text{COM}_x) \\ 
        & \quad - \ddot{y}_1 \, m_1 (x_1 - \text{COM}_x) - \ddot{y}_2 \, m_2 (x_2 - \text{COM}_x) 
   \end{aligned}
\end{equation} 
\end{figure} 

\section{Methodology} \label{sec:methodology}

\begin{figure*}
    \centering
    \centering \includegraphics[width=.9\hsize, page=2]{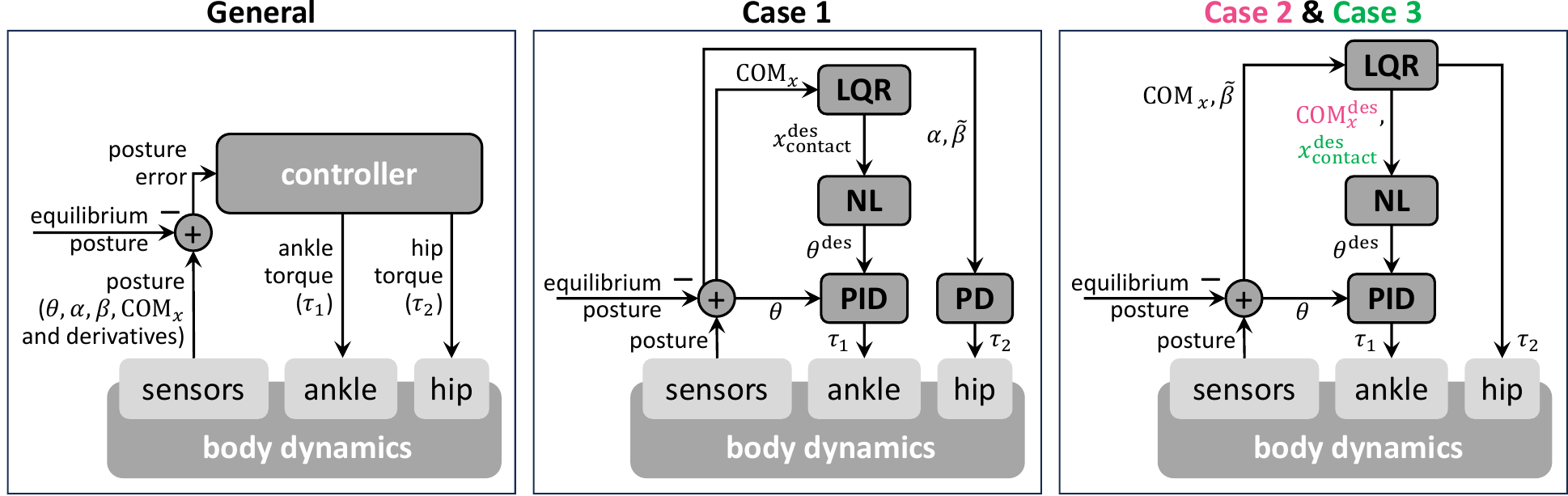}
    \caption{For visual simplicity, we omit derivatives from the diagrams; wherever a state is used (e.g., $\theta$), assume that its derivative (e.g., $\dot{\theta}$) is also used. \textbf{(Left)} General control strategy. \textbf{(Center)} Control strategy for Case 1: ankle and hip controllers are decoupled. \textbf{(Right)} Control strategy for Cases 2 and 3: the upper-layer LQR controller produces hip torque and desired states ($\text{COM}^\text{des}_x$ for Case 2 and $x^\text{des}_\text{contact}$ for Case 3) for the ankle. We apply a nonlinear transformation (`NL') on the desired states to obtain the desired foot angle $\theta$; this is tracked by a lower-layer PID controller which outputs ankle torque.} \vspace{-1.5em} \label{fig:diagram_control}
\end{figure*}

\begin{figure}
    \centering
\includegraphics[width=.8\hsize, page=3]{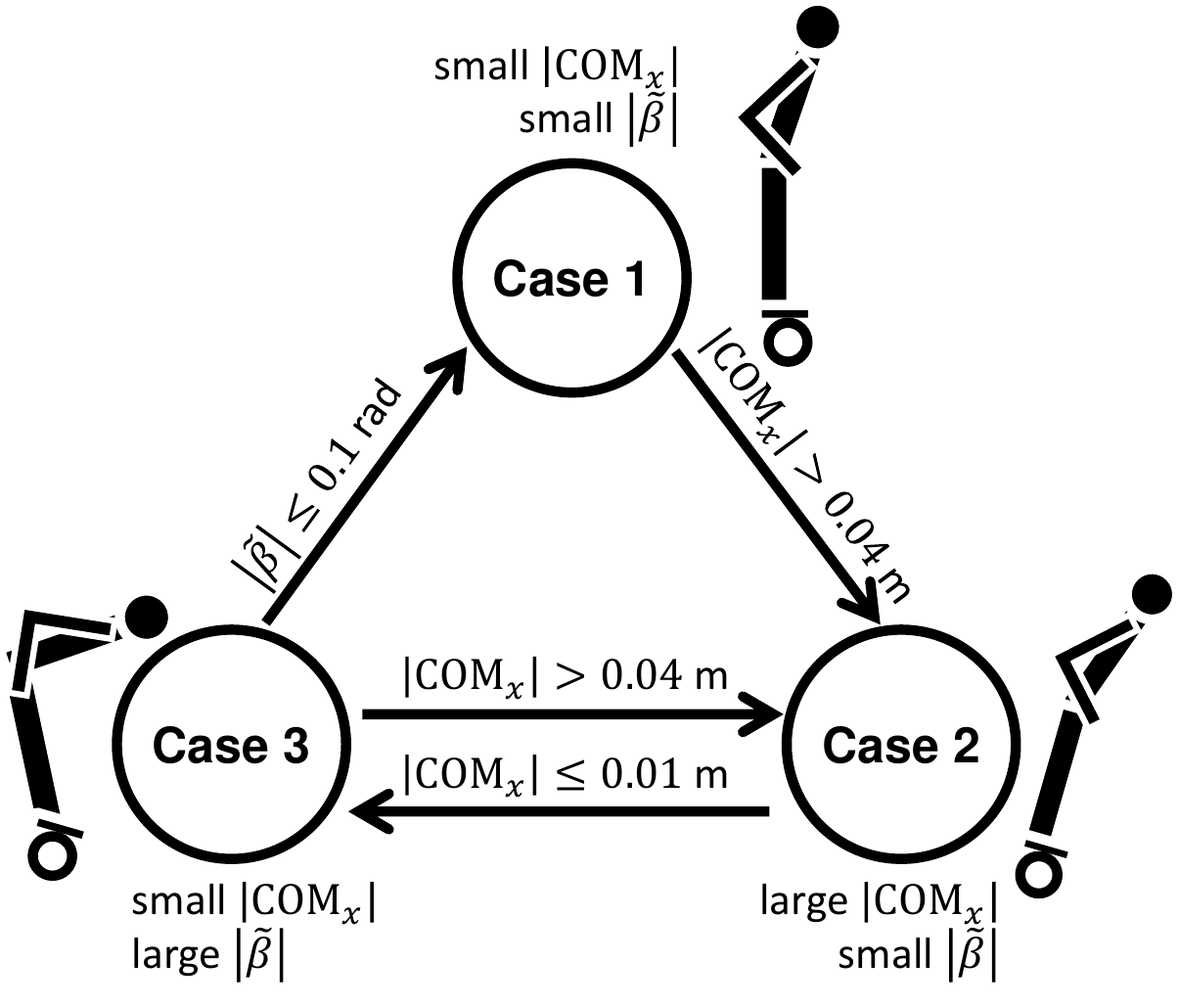}
    \caption{Switched controller: cases and transition rules.}
    \label{fig:diagram_cases}
\end{figure}

We found in initial experiments that a single offline controller (e.g. LQR or PID) only works well for a tiny portion of the state space. Thus, we partitioned the state space into three cases (see Figure \ref{fig:diagram_cases}, and developed one controller for each case (see Figure \ref{fig:diagram_control}). In Case 1, the COM and torso angle are close to equilibrium, and the controller only need to maintain the current state. In Case 2, the COM is far from equilibrium; the person is in immediate danger of falling, and the controller must aggressively and quickly reduce COM. In Case 3, the COM is close to equilibrium, while the torso angle is far from equilibrium; the person is not in immediate danger of falling, but the torso must be restored to equilibrium --- the controller must do so while keeping the COM close to equilibrium. In the absence of disturbances, we spend most of our time in Case 1; when a disturbance hits, we typically transition to Case 2, then to Case 3, and back to Case 1 (see Section \ref{sec:simulations}). The switching thresholds for torso angle and COM (see Figure \ref{fig:diagram_cases}) were selected via tuning during the controller design process based on the  capabilities of the individual controllers and general behavioral observations. For example, in the Case 1-to-2 transition, a threshold that is too small would cause the human to enter Case 2 unnecessarily often, expending extra energy on aggressive hip movements; however, a threshold that is too large would mean that we enter Case 2 only when the COM is very high, meaning that it may already be too late for Case 2 to stabilize the COM.

Even when partitioned into three cases, the strong nonlinearity of the dynamics makes it difficult to control the system with only an LQR or PID controller (per case). Thus, we combined these techniques together in a layered architecture for each case. The resulting controllers are simple to design and easy to understand and assemble, though they are tricky to provide theoretical guarantees for. One common theme across all three cases is that the foot angle ($\theta$) dynamics are complicated and nonlinear, which make linearization-based approaches inapplicable except for when $\theta$ is extremely small. Instead of working with $\theta$ directly in the model, we use other quantities (e.g. $\text{COM}_x$) that better lend themselves to linear assumptions, and apply a nonlinear transform to obtain a desired foot angle that is tracked by some lower-layer controller; this is detailed below.

\subsection{Case 1}
\label{case 1}
The Case 1 controller (see Figure \ref{fig:diagram_control}) is applied when both COM and torso angle are close to equilibrium. This controller keeps these values close to equilibrium. The hip and ankle are governed by separate controllers with no information exchanged. At the hip, we apply a high-gain PD controller that makes the hip angle $\eta := \beta-\alpha$ stiff:
\begin{equation}
    \tau_2=-K_p\eta-K_d\dot{\eta} 
\end{equation}

Since the hip is stiff, we can model the body dynamics as a single inverted pendulum with length $L$ with moving contact point $x_\text{contact}$ along the $x$-axis. The equation of motion is:
\begin{equation}
\frac{d^2\text{COM}_x}{dt^2}
    =g(\text{COM}_x-x_\text{contact}) /L
\label{eq:eom_case1_1}
\end{equation}
where $g$ is the gravitational constant. We can treat this as a linear state-space model with $\text{COM}_x$ and its derivative as state, and $x_\text{contact}$ as input. We apply an LQR controller\footnote{In simulation, we found that state penalty values on the order of 100 times the input penalty values was generally effective}. The LQR controller outputs the desired (i.e., optimal) values of $x_\text{contact}$, which we will call $x^\text{des}_\text{contact}$. We relate the desired contact point $x^\text{des}_\text{contact}$ to the desired foot angle $\theta^\text{des}$:
\begin{equation} \label{eq:case13_theta_xcontact}
    \theta^\text{des} = \sin^{-1} (x^\text{des}_\text{contact}/r)
\end{equation}
We apply a PID controller on the ankle that tracks $\theta^\text{des}$ and outputs the appropriate ankle torque $\tau_1$. 

\subsection{Case 2}
The Case 2 controller (see Figure \ref{fig:diagram_control}) is applied when the COM is far from equilibrium. This controller quickly reduces the COM to prevent a fall. We start with a state-space model: $\text{COM}_x$ and its derivative are states, and hip torque $\tau_2$ is input. In this case, since the controller acts very quickly, we can ignore the influence of gravity (and therefore of $\theta$). Then, our equations of motion \eqref{eq:eom_torque} can be written as \eqref{eq:case2_approx1}.

\begin{figure}
\begin{equation} \label{eq:case2_approx1}
\begin{aligned}
        F_x&=\ddot{x}_2m_2+\ddot{x}_1m_1, \quad F_y =\ddot{y}_2m_2+\ddot{y}_1m_1\\
        -\frac{\tau_2}{l_2}&=\ddot{x}_2m_2\cos{\beta}+\ddot{y}_2\sin{\beta}\\
        \ddot{\beta}l_2^2\frac{m_1m_2}{m_1+m_2}&=F_x(l_1\cos \alpha+l_2\frac{m_2}{m_1+m_2}\cos \beta)\\
        &\quad+F_y(l_1\sin \alpha+l_2\frac{m_2}{m_1+m_2}\sin \beta)\\
        \text{COM}_x &= \frac{x_1 \, m_1 + x_2 \, m_2}{m_0 + m_1 + m_2} \\
\end{aligned}
\end{equation}
\end{figure}

For $\alpha$ and $\beta$ both small\footnote{Since $\beta_0$ is small, $\beta$ being small is equivalent to $\tilde{\beta}$ being small}, we have:
\begin{equation} \label{eq:case2_1}
    \begin{aligned}
        \ddot{\beta}l_2^2\frac{m_1m_2}{m_1+m_2}&=F_x(l_1+l_2\frac{m_2}{m_1+m_2})\\
        F_x &= -\ddot{\alpha}l_1m_2-\ddot{\beta}l_2m_2-\ddot{\alpha}l_1m_1\\
        \tau_2 &= \ddot{\alpha}l_1l_2m_2+\ddot{\beta}l_2^2m_2\\
        \text{COM}_x&=-\frac{\alpha l_1 m_1+ \alpha l_1 m_2+\beta l_2 m_2}{m_0+m_1+m_2}
    \end{aligned}
\end{equation}

Rearranging, we see that:
\begin{equation} \label{eq:case2_2}
    \begin{aligned}
        \ddot{\alpha} &= \frac{\tau_2 (-l_1 - l_2)}{l_1^2 l_2 m_1} \\
        \ddot{\beta} &= \frac{l_1 \ddot{\alpha} (-l_1 m_1 - l_1 m_2 - l_2 m_2)}{l_2 m_2 (l_1 + l_2)}
    \end{aligned}
\end{equation}
From \eqref{eq:case2_1} and \eqref{eq:case2_2}, we see that $\frac{d^2\text{COM}_x}{dt^2}$ is proportional to $\ddot{\alpha}$ and $\ddot{\beta}$, and therefore proportional to $\tau_2$. Thus, the equations of motion can be approximated by the linear equation:
\begin{equation}
    \frac{d^2\text{COM}_x}{dt^2} = C_1 \tau_2
\end{equation}
where $C_1$ is some constant (found via \eqref{eq:case2_1} and \eqref{eq:case2_2}). We then apply LQR to obtain an optimal controller. In this case, we use high state penalty values (on the order of $10^8$ times the input penalty values) to facilitate aggressive hip action. The LQR controller produces optimal values of hip torque $\tau_2$ which are directly applied to the hip. 

To compute ankle torque $\tau_1$, we treat the closed-loop optimal state $\text{COM}_x$ as a desired center of mass, $\text{COM}^\text{des}_x$, and relate this to the desired foot angle $\theta^\text{des}$:

\begin{equation} \label{eq:ankle_nl}
\begin{aligned}
        \gamma &= -\sin^{-1}(\frac{\text{COM}^\text{des}_x}{r}) + \left\{
    \begin{array}{ll}
        -C_2, & \text{if } \text{COM}^\text{des}_x > 0 \\
        C_2, & \text{if } \text{COM}^\text{des}_x \leq 0 \end{array}
\right\} \\
        \theta^\text{des} &= \text{clip}(\gamma, \left[-\frac{\pi}{6}, \frac{\pi}{6}\right])
\end{aligned}
\end{equation}
Here, $C_2$ is a small positive constant, which is tuned to ensure foot angle smoothness when switching from Case 2 to Case 3; in particular, Case 2 features aggressive hip movement, while Case 3 features slow hip movement --- switching between these can cause sudden accelerations and discontinuities in the desired foot angle if $C_2$ is not tuned. Notice that the desired foot angle is clipped to ensure a reasonable range of motion; if the angle is not clipped, it is possible that the foot can be almost vertical, which is unrealistic. Finally, we apply a PID controller on the ankle that tracks $\theta^\text{des}$ and outputs the appropriate ankle torque $\tau_1$. 

\subsection{Case 3}
The Case 3 controller (see Figure \ref{fig:diagram_control}) is applied when the COM is close to equilibrium but the torso angle is far from equilibrium. This controller slowly restores the torso angle while keeping the COM close to equilibrium. We start with a state-space model with $\tilde{\beta}$, $\text{COM}_x$, and their derivatives as the states. The inputs are hip torque $\tau_2$ and $x_\text{contact}$. We use similar assumptions as Case 2; the equations of motion \eqref{eq:eom_torque} can again be written as \eqref{eq:case2_1} and \eqref{eq:case2_2}. For the influence of $x_\text{contact}$, we use similar derivations as Case 1, except now we have a double pendulum instead of a single pendulum. The equations of motion can be approximated by:
\begin{equation}
\begin{aligned}
    \ddot{\tilde{\beta}} &= C_3 \tau_2 \\
    \frac{d^2\text{COM}_x}{dt^2} &=g(\text{COM}_x - x_\text{contact})/L + C_4 \tau_2
\end{aligned}
\end{equation}
where $C_3$ and $C_4$ are constants (found via \eqref{eq:case2_1} and \eqref{eq:case2_2}). We then apply LQR to obtain an optimal controller. In simulation, we saw that overly aggressive correction of torso angle leads to instability; thus, we use state penalty values on the order of 10-100 times the input penalty values. The LQR controller produces optimal values of hip torque $\tau_2$ which are directly applied to the hip, as well as desired (i.e., optimal) values of values of $x_\text{contact}$, which we call $x^\text{des}_\text{contact}$. We relate the desired contact point $x^\text{des}_\text{contact}$ to the desired foot angle $\theta^\text{des}$ using \ref{eq:case13_theta_xcontact}. Finally, we apply a PID controller on the ankle that tracks $\theta^\text{des}$ and outputs the appropriate ankle torque $\tau_1$. 

\section{Muscle controller} \label{sec:muscle_controller}

While torque actuators are common in robotics (and make for easier math), humans actuate their bodies using muscle forces rather than joint torques. In this section, we briefly describe how to design a muscle controller using similar principles as the previous section. We focus on Case 1, and defer muscle control of the two other cases to future work.

First, we must build a plant model that includes muscles. A simplified model of the muscles involved in hip and ankle movement is shown in Figure \ref{fig:model_main}. Each joint is actuated by a pair of muscles, called an agonist-antagonist pair; in our model, muscles 1 and 2 actuate the ankle, while muscles 3 and 4 actuate the hip. Generally, one muscle in the pair contracts while the other relaxes \cite{Kim2018-hz}. Muscles generate force by contraction; they cannot generate force through expansion. Thus, two muscles per joint are required for full actuation in our model. For example, muscle 3 can be relaxed to let the human bend forward at the hip, but it cannot produce force to ``push'' the human into a bend.

The nervous system sends \textit{muscle activation} to the muscles, which determines the output force of the muscles, which result in change of joint angle. In our model, we assume that the $i^\text{th}$ muscle force follows equations provided in \cite{doi:10.1073/pnas.2319313121}: 
\begin{equation} \label{eq:muscle_activation}
    F_i = a_iF_i^\text{max} + F_i^0    
\end{equation}
where $a_i$ is activation, $F_i$ is force, $F_i^\text{max}$ is the maximum force for the muscle, $F_i^0$ is some equilibrium value.

\begin{figure}
    \centering
\includegraphics[width=0.8\hsize, page=5]{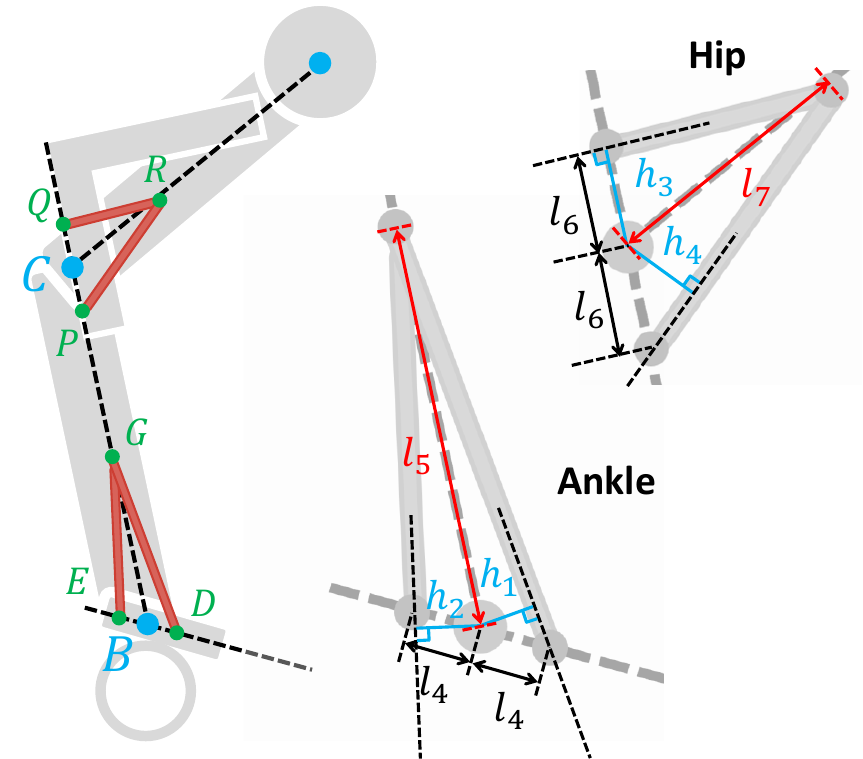}
    \caption{Geometric setup of simple muscle-based model.}
    \label{fig:muscle_geometry}
\end{figure}

To relate muscle forces to the equations of motion \eqref{eq:eom_torque}, we use techniques from \cite{li2004iterative}. For simplicity, we assume muscles are one-directional force actuators, and ignore the effects of length, velocity, and temperature. The setup is shown in Figure \ref{fig:muscle_geometry}; here, $l_4$, $l_5$, $l_6$, $l_7$ are constants related to muscle insertion locations, and $B, C, D, E, G, P, Q, R$ are written as coordinates, i.e., $B = (B_x, B_y)$. Let $M$ be the midpoint of the foot; torques and muscle forces are related via \eqref{eq:eom_muscle}.

\begin{figure}
\begin{equation} \label{eq:eom_muscle}
\begin{aligned}
    M &= (-r\sin\theta, \quad r\cos\theta) \\
    B &= (M_x - (l_0-r\theta)\cos\theta, \quad M_y - (l_0-r\theta)\sin\theta) \\
    D &= (B_x + l_4\cos\theta, \quad B_y + l_4 \sin \theta) \\
    E &= (B_x - l_4\cos\theta, \quad l_4\sin\theta) \\
    G &= (B_x - l_5\sin\alpha, \quad B_y + l_5\cos\alpha) \\
    C &= (B_x - l_1\sin\alpha, \quad B_y + l_1\cos\alpha) \\
    P &= (C_x + l_6\sin\alpha, \quad C_y - l_6\cos\alpha) \\
    Q &= (C_x - l_6\sin\alpha, \quad C_y + l_6\cos\alpha) \\
    R &= (C_x + l_7\cos\beta, \quad C_y +l_7\sin\beta) \\
    s_a &= l_4 l_5 \sin(0.5\pi - \theta + \alpha) \\
    \quad h_1 &= s_a\|F-D\|_2^{-1}, \quad h_2 = s_a\|F-E\|_2^{-1} \\
    s_h &= l_6 l_7 \sin(\pi - \alpha - \beta) \\
    h_3 &= s_h\|R-Q\|_2^{-1}, \quad h_4 = s_h\|P-R\|_2^{-1} \\
    \tau_1 &= F_2h_2-F_1h_1, \quad \tau_2 = F_4h_4-F_3h_3    
\end{aligned}    
\end{equation}
\end{figure}

\begin{figure}
    \centering
\includegraphics[width=0.65\hsize, page=4]{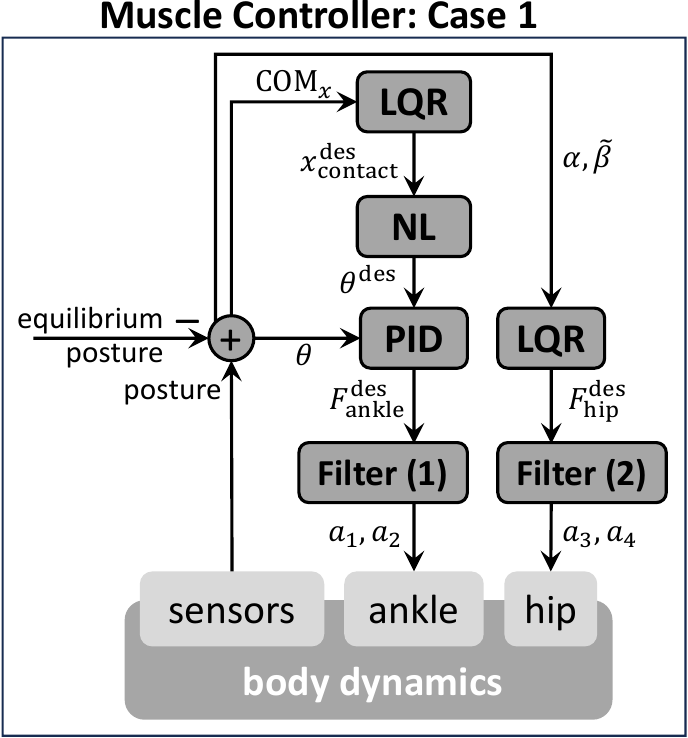}
    \caption{Control strategy for muscle-based actuation. `NL' represents a nonlinear transformation that takes the desired contact point and gives a desired angle. PID and LQR outputs undergo a filter that ensures smoothness, positivity, and appropriate scaling of the final muscle activation $a_i$; the specific filter equations vary slightly between the two. For visual simplicity, we omit derivatives from the diagrams; wherever a state is used (e.g., $\theta$), assume that its derivative (e.g., $\dot{\theta}$) is also used.}
    \label{fig:muscle_controller}
\end{figure}

The overall plant model becomes \eqref{eq:eom_torque}, \eqref{eq:eom_muscle}, and \eqref{eq:muscle_activation}; the control inputs are now muscle activation $a_i$ for $i=1,2,3,4$. As before, the controller has access to angles, angular velocities, and $\text{COM}_x$. To design the controller, we use a similar multi-layered approach as the torque controller (see Figure \ref{fig:muscle_controller}). One key difference --- other than the equations of motion --- is that activations are more restrictive than torques. Activations cannot be negative; this is related to the fact that a muscle can only generate force via contraction, not expansion. We will also normalize activations (i.e., $a_i \in [0,1]$) as is commonly done in the literature \cite{doi:10.1073/pnas.2319313121}.

As with the Case 1 torque controller, the hip and ankle are governed by separate controllers with no information exchange. The equations of motion for the hip are \eqref{eq:case2_2} and \eqref{eq:eom_muscle}; here, $F_3$ and $F_4$ are the control inputs to the plant. We linearize these equations, and apply an LQR controller to make the hip stiff, similar to the torque controller for Case 1. The LQR controller outputs the desired hip forces $F_3^\text{des}$ and $F_4^\text{des}$. We then apply a filter \eqref{eq:lqr_filter} to translate this into positive, normalized activation $a_i$ for $i=3,4$.
\begin{equation} \label{eq:lqr_filter}
    a_i = \max \left\{\ln(1+e^{F^\text{des}_i})/F_i^\text{max}, \quad 1 \right\}
\end{equation}

For the ankle controller, the first steps of the derivation are identical to the torque controller. We treat the system as a single inverted pendulum, and the equations of motion are \eqref{eq:eom_case1_1}. We apply the same LQR controller to produce the desired contact point $x_\text{contact}^\text{des}$; we then apply \eqref{eq:case13_theta_xcontact} to transform this into a desired foot angle $\theta^\text{des}$. We again use a PID controller to track this desired angle; however, this time, the controller outputs not torque but a desired scalar quantity $F^\text{des}_\text{ankle}$ representing the intended force at the ankle by some idealized bidirectional force actuator at the location of muscle 1. We then apply a filter \eqref{eq:pid_filter} on this quantity to obtain positive, normalized activation $a_i$ for $i=1,2$.
\begin{equation} \label{eq:pid_filter}
\begin{aligned}
    \text{if} \quad F^\text{des}_\text{ankle} > 0: &
    \quad  F_1^\text{des} = F^\text{des}_\text{ankle}, \quad F_2^\text{des} = 0 \\
    \text{if} \quad F^\text{des}_\text{ankle} \leq 0: &
    \quad  F_1^\text{des} = 0, \quad F_2^\text{des} = -F^\text{des}_\text{ankle} \\
    F_i^\text{des} &= \text{SmoothingFunction}(F_i^\text{des}) \\
    a_i &= \max \left\{ F_i^\text{des}/F_i^\text{max}, \quad 1\right\}
\end{aligned}
\end{equation}

\textit{Remark:} Another way of adapting the torque controller to muscle control is to build a conversion from torques to forces, which would be very likely to succeed. However, we did not take this approach because the goal of incorporating muscles is to increase the biological plausibility of the proposed model. A torque-to-force conversion is unlikely to be computed in the nervous system; humans use torques for mathematical simplicity, but there is no reason for the nervous system to compute using it.

\section{Simulations} \label{sec:simulations}

We now demonstrate the efficacy of our controllers in simulation. We use a person who is 1.6m tall ($l_0$ = 0.1m, $l_1$ = 0.8m, $l_2$ = 0.7m) and weighs 70 kg ($m_0$ = 15kg, $m_1$ = 35kg, $m_2$ = 20kg). The radius of the log is $r$ = 0.1m. The equilibrium torso angle is $\beta_0 = -0.5$ radians. We simulate the system for using an ODE solver (4th-order Runge-Kutta method) in MATLAB on the equations of motion.\footnote{We note that this method is numerically sensitive and requires very small time-steps (on the order of milliseconds) when $\text{COM}_x<0$.} Videos corresponding to Figures \ref{fig:simulation_3case}-\ref{fig:simulation_1case} can be found at \url{https://drive.google.com/drive/folders/10ZQUCWMf2cmncBXrUEBP6cFmdTj6hqD5}.

As a ``baseline'' comparison, we implemented a standard flat-ground posture controller, and found that it can't stabilize the system even when COM is close to equilibrium (i.e., $COM_x = 0.005$); that is, they do not work even in Case 1. This is likely due to two reasons: firstly, log-balancing gives rise to larger accelerations than flat-ground-balancing. Secondly, since the contact point between the foot and the log is not fixed, the direction (i.e., sign) of acceleration sign may change compared to the flat-ground-balancing case.

\subsection{Torque controller}

We conducted two sets of simulations for the torque controller. In the first set of simulations we explored the controller's ability to stabilize the system given some nonzero initial condition and zero disturbance. We found that as long as the initial value of $\text{COM}_x$ is within the range of $[-5, 9.6]$ cm, then our controller is able to stabilize (i.e., return to equilibrium). Most of the time, the controller starts in Case 2, goes to Case 3, then returns to Case 1.

\begin{figure}
    \centering
    \includegraphics[width=.88\hsize]{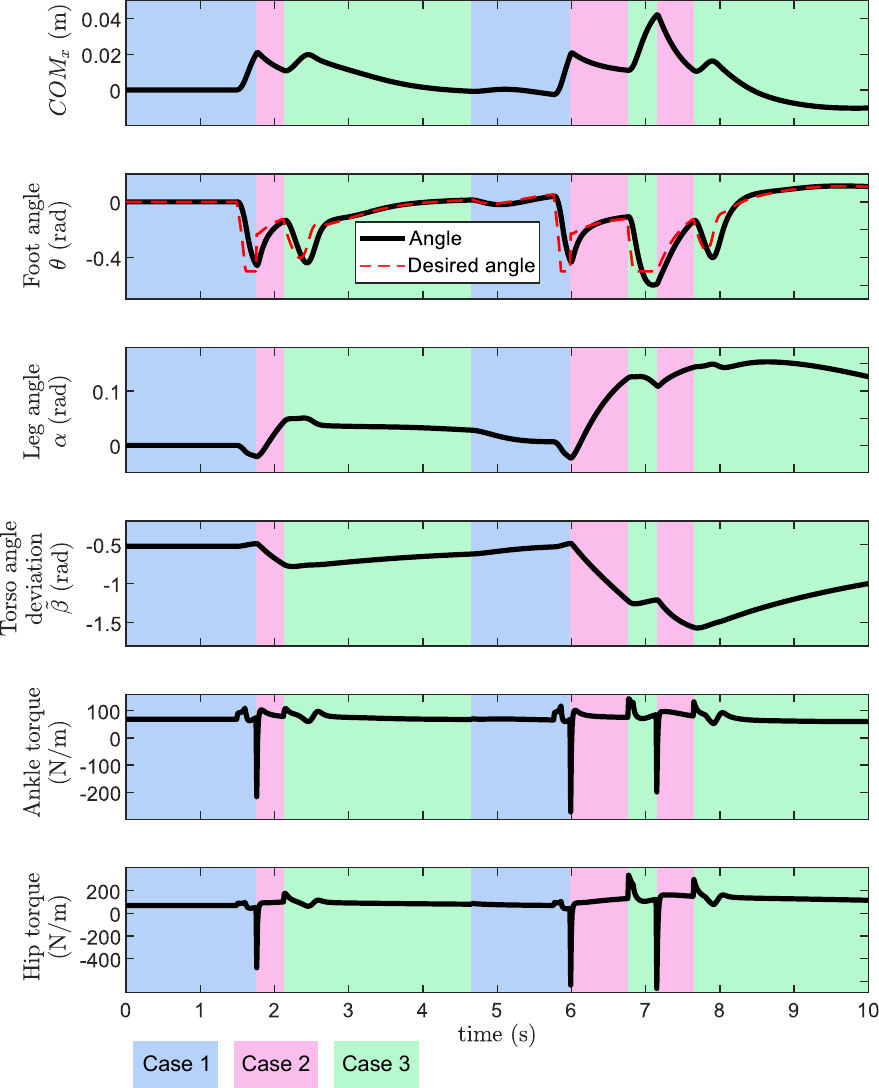}
    \caption{Trajectories of COM, body angles, and torques associated with the torque controller for a simulation with zero initial condition and three randomized kicks. Each kick is associated with a sharp change in torque. Background shading indicates which case the controller is in; in general, a kick puts the system into Case 2 (pink). After a short while (less than 1 second), the system enters Case 3 (green). The system eventually returns to Case 1 after 2-3 seconds. In this simulation, the system has not yet returned to Case 1 when the simulation finishes. Desired foot angle $\theta^\text{des}$ is shown by the dotted red line; generally, $\theta$ tracks this quantity quite well.}
    \label{fig:simulation_3case}
\end{figure}

In the second set of simulations, we assumed zero initial condition but applied disturbance forces at the hip. We partitioned the first 8 seconds of simulation into intervals of 0.1 seconds, and randomly chose 3 intervals in which to apply disturbance. The magnitude of the disturbance is chosen from a normal distribution centered at 50N with standard deviation 80N. Out of 1000 simulations, 867 were stabilized and 133 were not. Trajectories of COM, body angle, and torque for one of the stabilized simulations is shown in Figure \ref{fig:simulation_3case}. We see that the controller enters Case 2 after a kick, then transitions to Case 3, then back to Case 1 (if time permits). We also note that for all three kicks, Case 2 lasts less than 1 second, showing that the Case 2 controller indeed acts aggressively. In contrast, Case 3 takes longer.

\subsection{Muscle controller}

\begin{figure}
    \centering
    \includegraphics[width=.89\hsize]{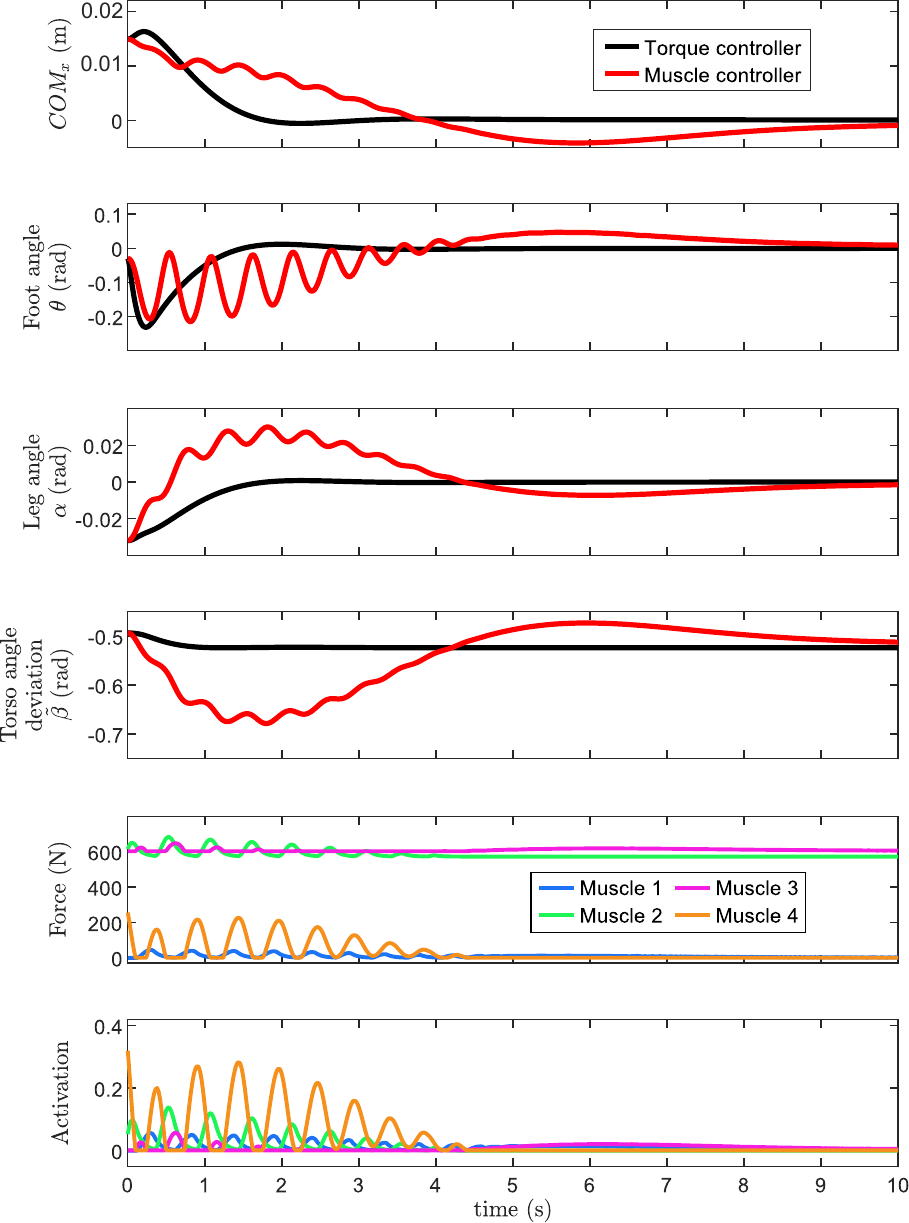}
    \caption{Trajectories of COM, body angles, muscle forces and muscle activations associated with the muscle controller for a simulation with small initial condition. COM and body angles for the torque controller are shown in black for comparison. The muscle controller exhibits oscillatory behavior, but does return the body to the equilibrium position after about 5 seconds.}
    \label{fig:simulation_1case}
\end{figure}

For muscle control, we use the following parameters: the maximum force of each muscle is $F_i^\text{max}$ = 800N; the equilibrium force of the muscles are $F_1^0$ = 0N, $F_2^0$ = 572N, $F_3^0$ = 602N, $F_4^0$ = 0N; the insertion lengths are $l_4$ = 0.15m, $l_5$ = 0.2m, $l_6$ = 0.2m, $l_7$ = 0.2m. In the zero-disturbance case, this controller is able to stabilize any initial condition that falls within Case 1 (and slightly beyond). Trajectories of COM, body angle, muscle activations and forces are shown in Figure \ref{fig:simulation_1case}. Compared to the torque controller, the muscle controller exhibits more oscillatory behavior; nonetheless, it restores the system to equilibrium after about 5 seconds. 

\section{Conclusions and future work} \label{sec:conclusions}

We proposed a torque- and muscle-based controller, and shown that each is able to stabilize the system. Future directions to explore include: (1) derivation of a stability proof for the switched controller, (2) extension of the muscle controller to all three cases, (3) inclusion of more realistic muscles (e.g. with dynamics and saturation), and (4) use of this model to capture real human balancing data for this task.

\bibliography{references}
\bibliographystyle{IEEEtran}

\end{document}